\documentclass[final,leqno,onefignum,onetabnum]{siamltex1213}

\usepackage{moreverb} \usepackage{amsmath}
\usepackage{tikz} \usepackage{pgfplots}

\newcommand\BibTeX{{\rmfamily B\kern-.05em \textsc{i\kern-.025em
b}\kern-.08em T\kern-.1667em\lower.7ex\hbox{E}\kern-.125emX}}

\def\mydelta{{ c }}
\newcommand{\HeatFlux}{\boldsymbol{\mathcal{Q}}}
\newcommand{\SpeciesFlux}{\boldsymbol{\mathcal{F}}}

\newcommand{\StressTensor}{{\tau}}
\newcommand{\ShearViscosity}{\eta}

\title{Achieving algorithmic resilience for temporal integration through spectral deferred corrections}

\author{R.W.Grout\footnotemark[2]
\and  H. Kolla\footnotemark[3]
\and  M.L. Minion \footnotemark[4]
\and J.B. Bell\footnotemark[4]}

\begin{document}
\maketitle

\renewcommand{\thefootnote}{\fnsymbol{footnote}}

\footnotetext[2]{Computational Science Center, National Renewable Energy Laboratory, Golden, CO 80401, USA} 
\footnotetext[3]{Scalable Modeling and Analysis Department, Sandia National Laboratories, Livermore, CA, USA}
\footnotetext[4]{Computational Research Division, Lawrence Berkeley
National Labs, Berkeley, CA 94720, USA}

\renewcommand{\thefootnote}{\arabic{footnote}}

\begin{abstract}
Spectral deferred corrections (SDC) is an iterative approach for constructing
higher-order accurate numerical approximations of ordinary
differential equations. SDC starts with an initial
approximation of the solution 
defined at a set of Gaussian or spectral collocation nodes
over a time interval and uses an iterative application of
lower-order time discretizations applied to a correction equation
to improve the solution at these nodes. 
Each deferred correction sweep 
increases the formal order of accuracy of the method 
up to the limit inherent in the accuracy defined by the 
collocation points.
In this paper, we demonstrate that SDC is well suited to recovering from soft
(transient) hardware faults in the data. A strategy where extra correction iterations
are used to recover from soft errors and provide algorithmic resilience is proposed. 
Specifically,  in this approach the iteration is continued until the residual (a
measure of the error in the approximation) is small relative to the
residual on the first correction iteration and changes slowly
between successive iterations.
We demonstrate the effectiveness of this strategy for
both canonical test problems and a comprehensive situation involving a
mature scientific application code that solves the reacting Navier-Stokes
equations for combustion research. 
\end{abstract}


\begin{AMS}\end{AMS}

\pagestyle{myheadings}
\thispagestyle{plain}
\markboth{SDC for algorithmic resilience}{Grout et al.}

\section{Introduction} 

Since its introduction by Dutt et al. \cite{DuttGR00}, the iterative
nature of spectral deferred corrections (SDC) has been leveraged
extensively to create efficient, high-accurate methods for temporal
integration tailored to specific types of problems.
For example, in Multi-Implicit Spectral Deferred Correction methods
\cite{bourliouxLaytonMinion:2003,LaytonM04}, 
the terms in an advection-diffusion-reaction system are integrated separately
with different timesteps but coupled together using the SDC approach to achieve
higher-order temporal accuracy than is achievable with traditional
operator-splitting schemes. A similar approach is used to reduce
splitting errors in a low-Mach combustion code by Nonaka et al.
\cite{NonakaBDGAM12}, where the SDC iterates are used to couple together interacting 
physical process. In this case, a significant advantage is realized in that the
reduction in splitting error reduces non-physical excursions into a chemical
state space that artificially excites the intrinsic stiffness in the system. SDC has also been used
to construct efficient time-parallel methods for partial differential equations (PDEs) \cite{EmmettM12}. Such desirable features that are not readily available in 
classical methods such as linear multistep or Runge-Kutta methods
can make SDC an attractive choice for time integration despite the 
fact that SDC often requires relatively more function evaluations per time step.

There is growing concern about the impact of hardware errors ---
especially those that can lead to successful completion with
erroneous results known as {\it silent data corruption}. This concern is
driven by trends towards increasing concurrency as well as
operation near design limits. Reducing voltage to improve energy
efficiency has long been known to increase susceptibility to soft
errors (e.g, \cite{Borkar99, DegalahalRVXI03}). Further, modern designs
tend to have elevated operating temperatures, which also increases the
soft error rate (\cite{SridharanL12}, \cite{Constantinescu02}). Wei et
al. define error resilience eloquently as \emph{the ability of a
  program to prevent an error from becoming a silent data
  corruption}. We will look to leverage the iterative nature of SDC to
provide algorithmic error resilience for temporal integration in the
face of soft errors in the arithmetic operations and scratch variables
used to update the solution.  We expect that the iterative nature
will be well suited to recover from transient errors. Chen et al. \cite{ChenBCP13} 
note that an adaptive RK scheme, where the solution update is
computed for two different timesteps, should be able to detect soft
faults, as the two evaluations will be dramatically different if a soft
fault has occurred. We use a similar logic when inspecting the
convergence rate between successive correction iterations to determine
if the solution is acceptable. 

The primary contributions of this paper are: firstly, to show that
monitoring the residual in SDC correction sweeps 
can be used to detect soft (transient) errors resulting from
hardware faults that could lead to silent data corruption using a
reference integration algorithm and secondly, to demonstrate the
feasibility of recovering from soft errors by continuing SDC correction
iterations.  The intent of this paper is not to look at the details of
low-level fault injection, but rather at how a time-integration
algorithm can recover from those faults that migrate up the call tree
through the
return values of kernels. Here we use the term
{\it kernel} to refer to routines at the application level
that compute terms in the governing differential equations being
integrated. For example, the kernels from the application discussed 
in Section~\ref{sec:s3d} are operations that compute advective or
diffusive terms for the method-of-lines
formulation or evaluate transport coefficients. 

The remainder of this paper is organized as follows: in the next
section, we present a brief outline of the SDC algorithm, relevant
aspects of the state of research on fault injection and algorithmic
resilience, and an overview of the combustion code used  as an
application benchmark later in the paper. We then
turn in Sect. \ref{sec:softerrinj}
to the behavior of the application in the context of
single occurrence synthetic errors, using the explicit Runge-Kutta
integrator traditionally employed in the application code as a baseline to
assess susceptibility to silent data corruption. We
also examine the ability of the SDC algorithm to recover from such
errors in an application test case and in a linear problem
to demonstrate how the damping proceeds in a controlled setting. Finally, 
in Sect. \ref{sec:multiple}, we
look at a comprehensive error injection test case where we inject
errors at an elevated rate into many runs of the application test case
to see how our SDC iteration strategy narrows the distribution of the simulation output in a
challenging scenario.

\section{Preliminaries and Related Work}

\subsection{SDC formulation} \vspace{-2pt}

Spectral deferred correction schemes were proposed by Dutt et
al. \cite{DuttGR00} and subsequently developed significantly by Minion and
colleagues (e.g., \cite{Minion03,bourliouxLaytonMinion:2003,LaytonM05}). 
The basic approach is
briefly recapped in this section before we consider additional aspects
of its performance relevant to use in practical applications.

SDC schemes are based on recasting the ordinary differential equation (ODE)
\begin{equation}
  \label{eq:7} 
  \phi' = F(\phi,t)  \qquad  \phi(t_n) = \phi_n
\end{equation}
over the time interval
 $ t^n \le t \le  t^{n+1}$
in integral form as
\begin{equation}
\label{eq:int}
\phi(t) = \phi_n + \int_{t_n}^t F(\phi,\tau)d\tau.
\end{equation}
Subdividing the interval $[t_n,t_{n+1}]$ by choosing 
$M+1$  Gauss-Lobatto quadrature 
nodes $t_m$ ($t^n = t_0 $ and $t^{n+1} = t_{M}$), for each node
we can write the approximation 
\begin{equation}
  \phi_m = \phi_n + \Delta t \sum_{j=0}^M q_{m,j} F(\phi_j,t_j).
\end{equation}
This integral approximation provides an accurate representation of the solution
$\phi_M=\phi_{n+1}$ at $t^{n+1}$; however,
it effectively couples the solution at all of the quadrature nodes in the interval. 
SDC can be thought of as providing an efficient iterative approach 
for computing the solution to this coupled system by iterative sub-stepping
over the nodes.

The basic idea is, given an approximate continuous solution ${\phi^k(t)}$,
one can define a residual that measures the error in the approximation $\phi^k$ as
  \begin{equation}
  \label{eq:residual} R( {\phi^k},t) = \phi_n +
\int_{t_n}^t F({\phi^k}(\tau),\tau) d\tau- {\phi^k(t)}.
\end{equation}
If we define $\mydelta^k (t) = \phi(t) -\phi^k(t)$, then by substituting the definition of the
residual into the integral form of the original equation we obtain
\begin{equation}
  \label{eq:correction} \mydelta^k(t) = \int_{t^n}^t \left[
F({\phi^k}(\tau) + \mydelta^k(\tau), \tau) - F({\phi^k}(\tau),
\tau) \right] d\tau + R({\phi^k}(t),t).
\end{equation}
We then discretize this equation using the approximate residual 
\begin{equation}
  \label{eq:residualdisc} R_m( \phi^k) = \phi_n +
\Delta t \sum^M_{m=0} q_{m,j} F(\phi^k_j, t_j) - \phi^k_m.
\end{equation}
An explicit Euler-type method to discretize
Equation~\ref{eq:correction} gives the resulting update formula for the
$k^\mathrm{th}$ iterate:
\begin{equation}
  \label{eq:discretedelta} \mydelta_{m+1}^k = \mydelta_m^k + \Delta t_m
\left[ F(\phi_m^{k+1}, t_m) - F(\phi_m^k, t_m) \right] +
R_{m+1}(\phi^k) - R_m(\phi^k),
\end{equation}
or, in a direct update form for $\phi_{m}^{k+1} = \phi_m^{k}+ \mydelta_m^k$,  
\begin{equation}
  \label{eq:8} \phi_{m+1}^{k+1} = \phi_m^{k+1} + \Delta t_m \left[
F(\phi_m^{k+1}, t_m) - F(\phi_m^k, t_m) \right] + I_m^{m+1}(\phi^k),
\end{equation}
 where:
\begin{equation}
  \label{eq:Iapprox} I_m^{m+1}(\phi^k) = \Delta t \sum_{j=1}^M (q_{m+1,j}-q_{m,j})
F(\phi^k(t_l), t_l) \approx \int_{t_m}^{t_{m+1}}
F(\phi^k(\tau),\tau)d\tau.
\end{equation}
$I_m^{m+1}$ is the equivalent to the integral of the polynomial interpolant
of $\phi^k$ over the interval $[t_m,t_{m+1}]$.
Each such iteration can improve by one the formal order
of accuracy of the approximate solution up to the order of the
underlying quadrature.  In the case of $M+1$ Lobatto nodes, the method
achieves order $2M$ on convergence.

\subsection{Soft error fault injection}
\label{sec:faultinj}

For the purpose of this study we follow the taxonomy of Hoemmen and Heroux \cite{HoemmenH11}, wherein
hard faults are those that cause program interruptions and clearly denote an incomplete program execution,
while soft faults are typically observed as 
random bit flips, where one or more bits of memory are reversed. These faults are
transient and do not indicate hardware damage, as opposed to persistent faults such as
bits that are immutable due to a physical defect (`stuck bit' errors).
Depending on where in the memory hierarchy they occur and the robustness of the algorithm, soft faults
may not always lead to a solution failure but might result in an erroneous solution 
despite completely evading detection \cite{ElliottMSW13}.
It might be acceptably inexpensive to provide soft fault detection and 
correction mechanisms for some, but not all, memory levels.
For instance, error correction codes have been shown to correct a majority of soft faults in main memory \cite{SridharanL12}, while processor registers 
are difficult to protect from soft faults \cite{HwangSS12}.   
Many factors such as altitude, age, temperature and utilization are thought to 
affect error rates in real machines with a significant variability observed across various DRAM vendors. 
Recent studies have attempted to characterize and quantify error rates 
by surveying hardware logs from real machines, although a consensus is far from apparent.
Schroeder et al. \cite{SchroederPW11} study error rates from commodity clusters in Google's server fleet
and observe that a majority of the errors are hard errors and soft errors 
are far less probable (a soft error probability of $\sim 2$\% for every hard error).
On the other hand Sridharan et al. \cite{SridharanSDBG13} 
find the opposite to be the case from a survey of data from two high-performance computing systems: Cielo at Los Alamos National Laboratory 
and Jaguar at Oak Ridge National Laboratory. Nonetheless, the most dominant mode seems to be single-bit errors (60\%)
with hard and soft errors being approximately equi-probable.

Considering the various enmeshed layers of software and hardware, the propagation of soft faults from one layer to  another can be complicated to model.
Strictly speaking, a bit flip at the level of 
hardware instructions is unlikely to migrate up to the application level as a single bit flip 
after several operations have been performed on the data. Even near the hardware level, a single bit flip in an instruction input might result 
in multiple bit flips in the destination register \cite{ElliottMSW13}. Despite this, there is some evidence that injecting
single bit flips at higher levels produces similar effects from an application
perspective as injecting errors near the hardware level. We choose this approach because
it allows us to reason about the algorithmic sensitivity to the errors while eliminating 
the potentially confounding effects of interaction of the errors before reaching the application
level. 

Wei et al. compare the behavior of high-level
fault injection (implemented at the LLVM Intermediate Representation (IR) level) to low-level
fault injection (using Intel PIN tools) and find that, while there
were significant differences in the number of program crashes
between the two techniques, the IR-level fault injection is
effective for assessing the impact of soft faults that
result in silent data corruption.  Wei et al. \cite{WeiTLP14} also note that 
it is an established {\it de facto} standard that single bit flips
\cite{FangPRG14} are an appropriate approach. In a related issue, Fang et al. \cite{FangPRG14} 
look at the effect of
  fault injection on multithreaded programs implemented using
  OpenMP and consider the sensitivity of the thread where the faults 
  are injected due to the emphasis of the master thread on problem
  setup/teardown (phases of their chosen benchmarks that are
  particularly prone to resulting in ultimate silent data corruption
  in the output from fault injection). In our present application
  of interest, the setup/teardown phases are a very small portion
  of the overall runtime, and otherwise the application follows a
  bulk-synchronous model.

Since our focus here is on the algorithmic robustness of SDC, we adopt a simple fault injection model. Considering that processor registers and 
arithmetic lookup units (ALUs) / floating point units (FPUs) are the most
vulnerable to soft faults \cite{WeiTLP14}, 
we model soft faults as single bit flips in processor registers. However, we inject errors at the level of the application
rather than at or very near the hardware level.  
We adopt an approach similar to, but even closer to the application level, than that of Wei et al. \cite{WeiTLP14} and inject faults as if they manifest as single bit flips in register work arrays of 
the application level kernels that evaluate the terms contributing to the time derivative ($F$) of our system of ordinary differential equations.

\subsection{Algorithmic approaches to resilience}
Since a large number of scientific applications employ linear system
solvers, methods to incorporate resilience in iterative linear solver
algorithms have received wide attention. For example, Hoemmen and Heroux
\cite{HoemmenH11} propose a {\it Fault-Tolerant} version of the
generalized minimal residual method (FT-GMRES) whereby the inner
iteration that corresponds to the preconditioning step for the outer
iteration is allowed to be unreliable. Rank deficiency of the
subsequent upper Hessenberg matrix could signal a potentially faulty
execution of the inner iteration that would require some recovery
strategy. The decision about whether a fault has occurred, and the
subsequent recovery, is a global operation and involves agreement and
hence global communication. Sloan et al. \cite{SloanKB13} suggest
that error detection and recovery should instead be localized near the
fault occurrence. The most expensive computational kernel in linear
solver algorithms such as GMRES, quasi-minimal residual method (QMR) and
conjugate gradient (CG) is usually a matrix-vector multiplication. Sloan
et al. \cite{SloanKB13} contend that a soft error is most probable
in this kernel and suggest an identity check that involves projecting
the result of the matrix-vector multiplication onto a test vector. The
projection can be computed two different ways, so the results should
agree if there were no faults in the original matrix-vector
multiplication. By choosing the test vector to initially have all
elements set to unity, they suggest a recursive hierarchical algorithm
to hone in on the exact locations of faulty execution. Stoyanov and
Webster \cite{StoyanovW13} consider Jacobi and Gauss-Siedel fixed point
iteration algorithms and leverage the identity that the norm of the
difference between successive iterates should reduce at the same rate as
the rate of convergence of the algorithm. They suggest that checking 
this identity can be used as a method to identify errors due to soft
faults and propose rejecting iterations that fail this test as a means
to incorporate resilience.

Alternative approaches have also been proposed for explicit PDE schemes
that are not iterative in nature. Mayo  et al. \cite{MayoAR12}
suggest combining two extremes in the tradeoff space for resilient
explicit PDE algorithms: {\it artificial viscosity}, the physical
mechanism that damps perturbations, and, {\it triple modular
redundancy}, the strategy of performing computations three times and
accepting a result that was reproducible at least twice.
They propose using multiple finite difference
schemes over stencils of different widths at each grid point of the same
formal order of accuracy to identify and discard outliers that might
have been corrupted due to soft faults. Donzis and Aditya
\cite{DonzisA14} propose asynchronous explicit finite difference schemes
for PDEs that could be viewed as a potential resilience strategy.
Typically, the explicit scheme for spatial derivatives requires the
solution from neighbouring grid points, which involves communication of
ghost regions across processing elements (PEs). In the conventional
implementation of such schemes the communication and the calculation of
spatial derivatives are completed by all PEs before the next timestep is
begun; i.e., all portions of the domain advance the solution in a
time-step synchronised fashion. However, one might envision that soft
faults cause some portions of the domain to take longer to execute
an iteration, introducing an asynchrony between PEs. Donzis and Aditya
\cite{DonzisA14} propose asynchronous schemes whereby neighboring PEs
could be at different timesteps but still perform spatial derivatives
to an intended order of accuracy. They model the asynchrony between
neighboring PEs as a random process and show that while such schemes
can be stable the accuracy in both time and space might be degraded.  
However, the desired order of accuracy severely limits the maximum asynchrony
allowable between any pair of PEs.

\subsection{S3D reacting flow solver and ignition benchmark
  problem}
\label{sec:s3d}
S3D is a solver for compressible reacting flows
developed by Chen et al. \cite{s3d}. S3D uses eighth-order
finite-difference approximations of  spatial derivatives
with a method-of-lines discretization integrated temporally using a
six-stage, fourth-order compact Runge-Kutta integrator from the family
developed by Kennedy and Carpenter \cite{KennedyCL00}. Second
derivatives are obtained by repeated application of the discrete first
derivative operator. The code has
been used to produce direct numerical simulations (e.g., sufficiently
resolved to capture all relevant continuum scales for turbulence,
chemical reaction and turbulence-chemistry interaction) of a variety
of turbulent combustion problems. Past problems include premixed
flames \cite{HawkesC05,SankaranHCLL07,GruberSHC10}, non-premixed flames
\cite{HawkesSSC07,YooSC09, GroutGYC11,GroutGKBBGC12} and autoignition
problems \cite{EchekkiC03,SankaranIHC05}. The code solves the
compressible Navier-Stokes equations along with transport of the mass
fractions of $K$ chemically reacting species using a mixture averaged
transport model.
The species density, momentum and energy equations of hydrodynamics are given by
\begin{equation}
\frac{\partial }{\partial t} \left( \rho_k \right)  + {\bf \nabla} \cdot \left( \rho_k {\bf v} \right) +
{\bf \nabla} \cdot {\SpeciesFlux}_k =  \dot{S}_k, 
\label{eqn:spec}
\end{equation}
\begin{equation}
\frac{\partial }{\partial t} \left( \rho {\bf v} \right)
+ {\bf \nabla} \cdot \left[ \rho {\bf v} {\bf v}^T +  p \mathbf{I} \right]
+ {\bf \nabla} \cdot  {\StressTensor}  = 0,
\label{eqn:mom}
\end{equation}
\begin{equation}
\frac{\partial }{\partial t} \left( \rho E \right)  + {\bf \nabla} \cdot \left[ (\rho E + p) {\bf v}  \right] +
{\bf \nabla} \cdot \left[ {\HeatFlux} + \StressTensor \cdot {\bf v} \right] = 0,
\label{eqn:energy}
\end{equation}
where $\rho_k$, ${\bf v}$, $p$,  $E$ and $\dot{S}_k$ denote, respectively, the mass density for species $k$,
fluid velocity, pressure,  total specific energy and chemical source term for species $k$
for a mixture with $K$ species
($k=1,\ldots K$).
We note that $\sum_k \SpeciesFlux_k = 0$ and $ \sum_k \dot{S}_k = 0$, so that summing the species equations
gives conservation of mass with $\sum \rho_k = \rho$, the total fluid density. 
Note that $\mathbf{v}\mathbf{v}^T$ is a (tensor) outer product with $T$ indicating transpose and
$\mathbf{I}$ is the identity tensor (i.e., ${\bf \nabla} \cdot p \mathbf{I} = {\bf \nabla} p$).
Transport properties are given in terms of
the species diffusion flux, ${\SpeciesFlux}$, viscous tensor, ${\StressTensor}$, and heat flux, ${\HeatFlux}$.
The viscous tensor is,
\begin{equation}
{\StressTensor} =
-\ShearViscosity \left ( \nabla \mathbf{v} + (\nabla \mathbf{v})^T  \right )
+\frac{2 }{3} \ShearViscosity  \left({\bf \nabla} \cdot {\bf v}\right) \mathbf{I},
\end{equation}
where $\ShearViscosity$ is the shear viscosity.
The heat flux is 
\begin{equation}
\HeatFlux = \sum_k h_k \SpeciesFlux_k -\lambda \nabla T,
\end{equation}
where $h_k$ is the enthalpy of the $k$ species and $\lambda$ is the thermal conductivity.

The diffusion velocity of the $k^{th}$ species is modeled
with a mixture-average formulation for $k-1$ species:
\begin{equation}
\SpeciesFlux_k 
 = - \overline{D}_k \left[ \nabla Y_k + Y_k(\overline{W}\nabla W_k + (1-M_k \overline{W}) \frac{1}{p}\nabla p \right ],
\end{equation}
where $Y_k = \rho_k/\rho$ is the mass fraction of species $k$, $\overline{D}_k $ is the mixture averaged
diffusion of species $k$, $W_k$ is molecular weight of species $k$ and
$\overline{W}$ is the mean molecular weight.
The final species diffusion velocity is computed so as to enforce conservation of mass:
\begin{equation}
	\SpeciesFlux_{K} = \sum_{k=1}^{K-1} - \SpeciesFlux_{k}, \;\;
\end{equation}
where $K$ is the dominant species, typically $N_2$.
Thermodynamic properties are temperature dependent; the temperature is related to the 
energy by:
\begin{equation}
    \label{eq:eofT}
E = e_s + \frac{1}{2} u_k u_k; \qquad e_s = \int_{T_0}^T C_vdT - \frac{ R T_0}{\overline{W}}.
\end{equation}
where $C_v$ is the mixture constant volume specific heat and $R$ is the ideal gas constant.

The chemical source terms appearing in the species equations are computed by evaluating 
a chemical reaction network 
\begin{equation}
    \dot{S}_k = W_k \sum_{j=1}^{N_r} \nu_{kj} R_j,
\end{equation}
where $\nu_{kj}$ are the stoichiometric coefficients for reaction $j$ 
and the rates of the $N_r$ reactions are given by expressions of the Arrhenius form used by \cite{KeeRMM96}.
For example, for a reaction where reactants $A$ and $B$ are converted into products $C$ and $D$:
\begin{equation}
A + B \Leftrightarrow C + D,
\end{equation}
the forward rate is given by:
\begin{equation}
R_f = [A][B] k_f; \qquad k_f = A_{fj}T^{\beta_j}\exp\left(\frac{-T_{aj}}{T} \right),
\end{equation}
where
$A_{fj},\beta_j, T_{aj}$ are coefficients
describing the $j^{th}$ reaction with the reverse rate given by:
\begin{equation}
    \label{eq:Rb}
	R_b = [C][D] k_b ; \quad k_b = \frac{k_f}{k_{eq}}; 
	\quad k_{eq} =\left(  \frac{p}{RT} \right)^{\sum_{n=1}^N \nu_{nj}} \exp \left( \frac{\Delta S_j^0}{R} - \frac{ \Delta H_j^0}{RT} \right)
\end{equation}
where $\Delta S_j^0,\Delta H_j^0$ are the entropy and enthalpy of formation
difference across the reaction, respectively. 

The ideal gas equation of state ($p=\rho RT/ \overline{W}$) completes the description of the system.
To solve Equations~\ref{eqn:spec}--\ref{eqn:energy}, a method-of-lines approach is used where
spatial derivatives are replaced by a finite-difference operator of the form:
\begin{equation}
	\left[ \frac{\partial \phi }{\partial x}\right]_i \approx \sum_{m=1}^{4} \left( \alpha_m \phi_{i-m} + \alpha_m \phi_{i+m} \right).
\end{equation}
In the course of evaluating the time derivatives, S3D computes the various terms 
much as written here where various kernels (e.g., compute operand, apply derivative
operator, compute diffusion velocity) operate on the entire solution grid until all
of the time derivatives are completely assembled. 
The resulting system of ODEs is then integrated with an explicit time
integration method. 
The standard time integration approach in S3D is a six-stage
fourth-order Runge-Kutta method from the family of 
of Runge-Kutta integration schemes proposed by Kennedy and
Carpenter~\cite{KennedyCL00}.

\begin{figure}[htb] \centering $\begin{array}{c}
  \includegraphics[width=0.8\columnwidth]{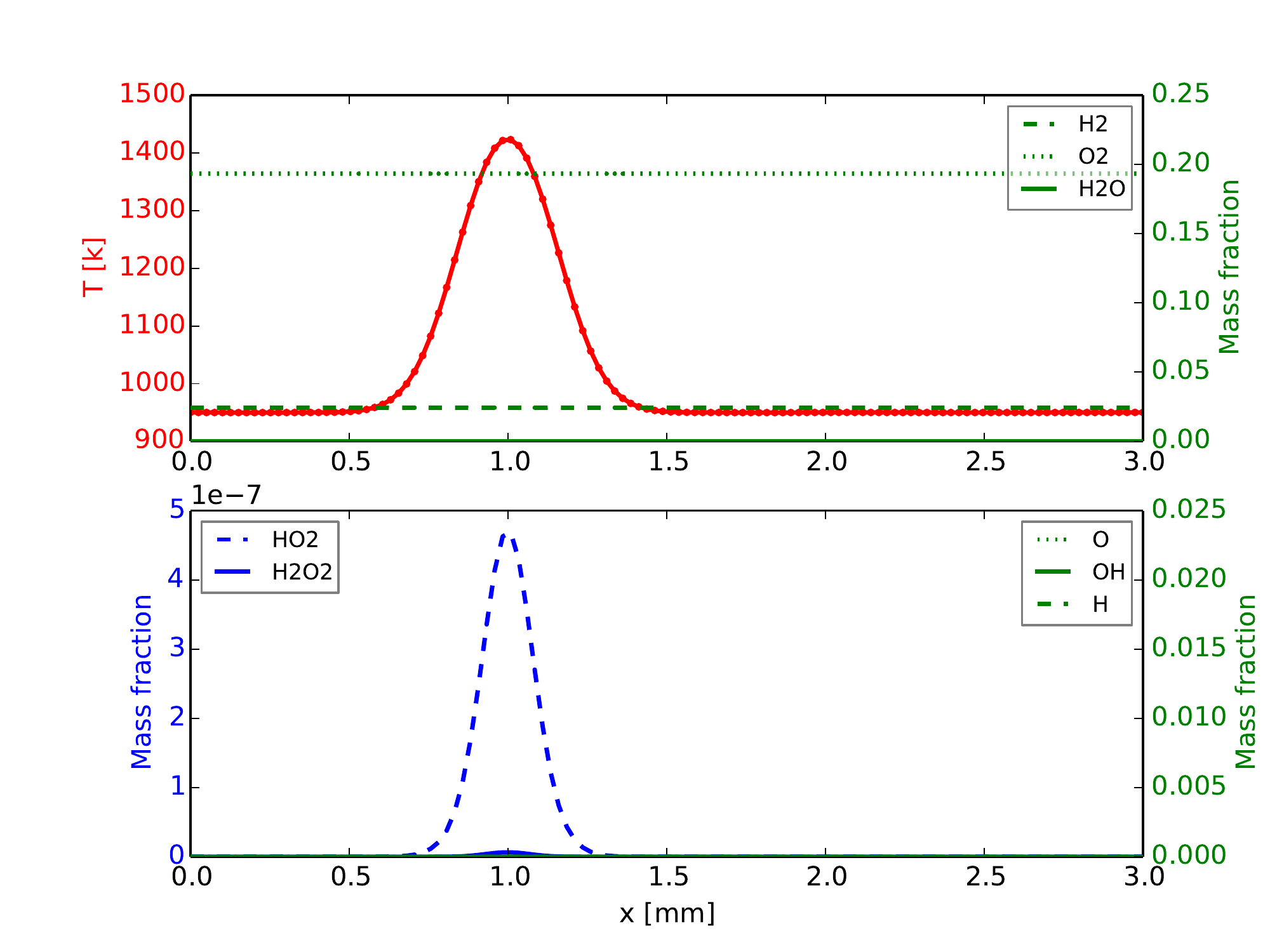}\\
  \includegraphics[width=0.8\columnwidth]{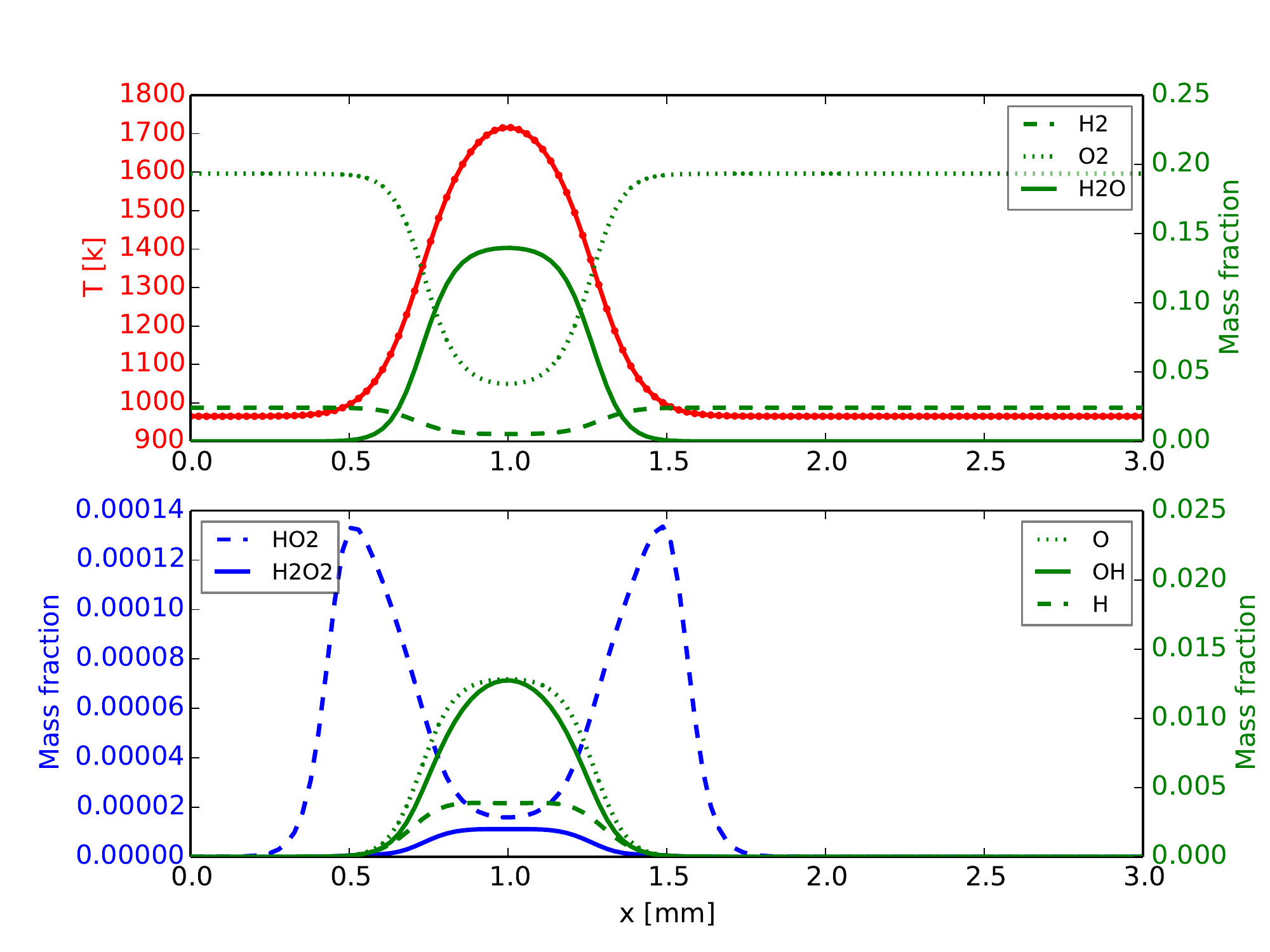} \\
\end{array}$
  \caption{Spatial profiles of temperature and species mass fractions
at $t=5.5\mu s$ (top), and $t=30\mu s$ (bottom) from reference solution obtained with
6,4-RK integrator}
  \label{fig:1dignsolp}
\end{figure}

\begin{figure}[htb] \centering
  \includegraphics[width=0.8\columnwidth]{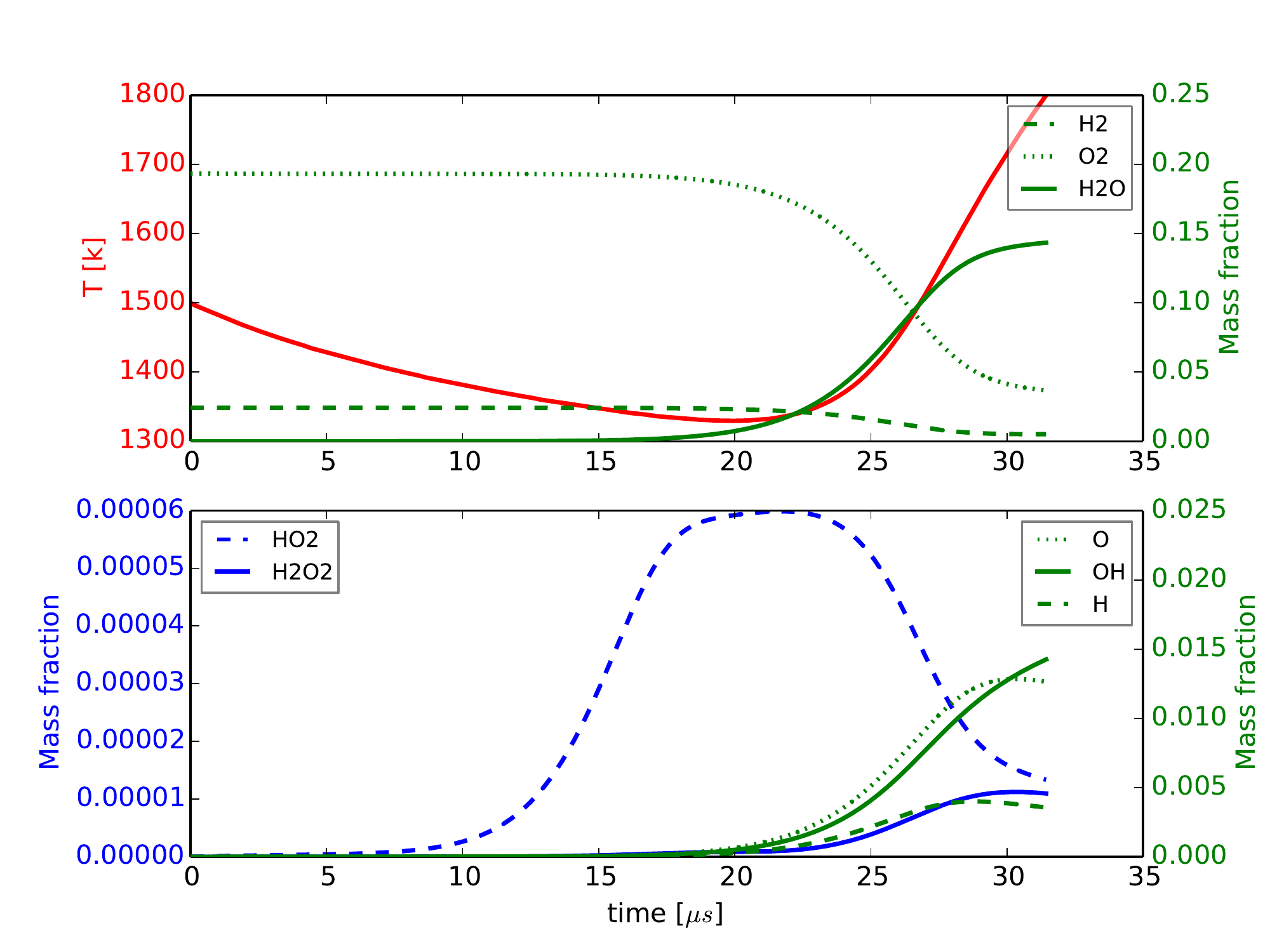}
  \caption{Temporal evolution of maximum temperature and species mass
fractions at gridpoint coinciding with maximum temperature from
reference solution obtained with 6,4-RK integrator}
  \label{fig:1dignmax}
\end{figure}

In the tests that follow we will use a fixed timestep and
tolerate the extra computational cost as a necessary expense to remove one aspect that
would make the results more difficult to interpret; in future work we plan to 
study the combination of SDC and adaptive time step control. The canonical problem is a
one-dimensional simulation of a homogeneous mixture composed of
hydrogen and air mixed in a stoichiometric ratio with a Gaussian
temperature hot spot placed in the center of the domain according
to:
\begin{equation} T(x) = T_0 + (T^*-T_0)\frac{1}{\sigma\sqrt{2\pi}}
e^{-(x-x^*)^2/(2\sigma^2)}.
\end{equation} Solutions for this problem obtained using S3D and
the native integrator used historically in S3D (the 6,4-RK algorithm) are
shown in Figure~\ref{fig:1dignsolp} and Figure~\ref{fig:1dignmax}. The
problem is one dimensional; 120 grid points are used to spatially resolve the ignition 
process and a fixed timestep of 5ns is used in all cases.  
Figure~\ref{fig:1dignsolp} shows that
the initial spatial temperature profile drives formation of a broad pool of
hot radicals, led by $HO_2$ that is eventually consumed as the mixture
proceeds towards ignition and takes the first steps towards the
formation of a front. 
In the time histories shown in Figure~\ref{fig:1dignmax}, the peak
temperature decreases due to diffusive processes along with the slow
buildup in $HO_2$ followed by an increased in $H_2O_2, OH, O$ and
finally a rapid rise in temperature. 
The chemical mechanism is that of Li et al. \cite{LiZKD04};
\emph{CHEMKIN's} \cite{KeeRMM96} \emph{tranlib} is used to evaluate
transport coefficients for a mixture-averaged diffusion formulation. This
test case has a relatively long `soaking' period, requiring approximately
5000 timesteps before the onset of thermal runaway at 20~$\mu s$. This
provides ample opportunity for small errors to compound into a large
effect on the solution yet is relatively manageable for experimentation. A
similar test case, a zero-dimensional ethylene-air ignition problem, is
used by Spafford et al. \cite{SpaffordMVCGS09} to study the effects of
single-precision on chemical reaction rate evaluation in the context of
porting S3D kernels to a graphics co-processor, where the test case proved
sufficiently sensitive to the accuracy of the function evaluation that
evaluating the reaction rates in single precision is insufficient to
achieve an acceptable solution. 


\section{Soft error injection response}
\label{sec:softerrinj}

In this section we look at injecting two types of soft errors into
major work arrays (those of the dimension of the solution grid) during
the computation of the solution: 
\begin{itemize}
\item[A.]  Scaling a single value within a work array by
a large factor (i.e., multiplying by $10^4$)
\item[B.] Reversing
the value of a bit at any position within the array (i.e., the value at any
gridpoint could have any bit within it flipped, including the sign
bit, the mantissa and the exponent positions).
\end{itemize}
We use the  type A errors 
to explore the sensitivity of the solution to various
intermediate values and to study how  continued SDC sweeps
can correct for such errors. 
Type A errors produce a moderate
response in that they typically produce a perturbed state that is incorrect but still
physically plausible---the circumstance where silent data corruption
is intuitively likely. Type B errors are more realistic, but can
result in perturbed states that are physically inconsistent
(e.g., negative temperatures). We use the bit flip approach,
described in Section ~\ref{sec:faultinj}, for a comprehensive
assessment of the technique integrated into the application code at
the end of this section. In all cases, we limit our study to the work
arrays that form return values of basic ``simulation kernels'' (which will be described in
the following section). In other words,
we leave persistent variables (e.g., stencil coefficients),
control flow and instruction logic, and the solution vector at the
start of the timestep unperturbed.

\subsection{Work array sensitivity}
\label{sec:sens}
\begin{figure}[htbp] \centering
  \includegraphics[width=0.8\columnwidth]{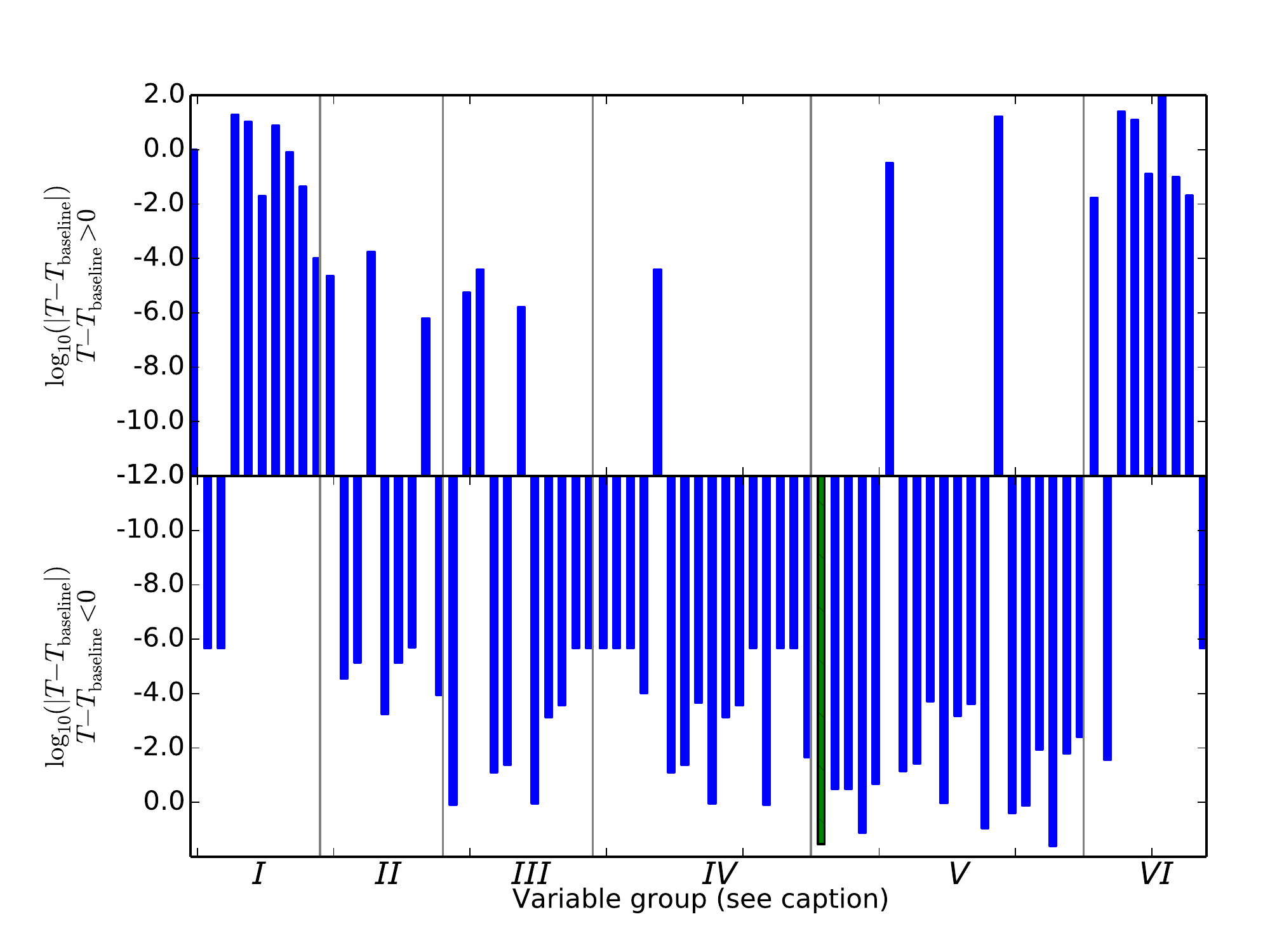}
  \caption{Difference in maximum temperature in domain from baseline
    at fixed time near end of ignition delay resulting from one-time
perturbation of work-arrays during calculation using traditional
(6,4-RK) integration algorithm. Derivative of density
is highlighted in green crosshatch. Work arrays where perturbation resulted in
simulation crash are not shown. Variable groups are as follows: Group I, primaries ($u,  \gamma, c_p, Y_\alpha$); 
Group II, enthalpies ($h_\alpha$); Group III, 
gradients ($\nabla u, \nabla T, \nabla Y_\alpha$); Group IV, diffusive fluxes ($\tau_{ij},J_\alpha,J_T$);
Group V, second derivative operands, results (momentum, energy, species); Group VI, reactions ($S_\alpha$).}
  \label{fig:senscope}
\end{figure}

The S3D algorithm computes several quantities
that are stored in work arrays during the evaluation of the right-hand
side function, and the sensitivity of the solution to  
perturbations varies widely between quantities.  
To demonstrate this, we modified the code that evaluates the 
temporal derivatives of given quantities
so that the results of kernel functions are perturbed. That
is, evaluation of the time derivative involves application of several
kernels called as functions that manipulate a set of work arrays:
\begin{equation}
  \label{eq:12}
  \mathcal{R}^k(\vec{W}) \gets \mathrm{\sc kernel}_k
  (\mathcal{I}^k(\vec{W}), \vec{q}, \dots),
\end{equation}
where $\vec{W}$ is the vector of multidimensional work arrays, $\mathcal{R}^k(\vec{W}) $ is the subset of the work arrays
altered by the $k^\mathrm{th}$ kernel, $\mathcal{I}^k(\vec{W})$ is
the subset of the work arrays used as input to the $k^\mathrm{th}$ kernel,
$\vec{q}$ is the vector of conservative state variables at the start
of the timestep and the $(\dots)$ represents the constants that complete
the closure for the kernel. In this nomenclature we apply a
perturbation function $\mathcal{P}$ that applies a single bit error (as discussed near the end of
Section~\ref{sec:faultinj}) to the return values of each kernel
immediately after each kernel completes:
\begin{equation}
  \label{eq:13}
  \mathcal{R}^k(\vec{W}) \gets \mathcal{P}[\mathcal{R}^k(\vec{W})].
\end{equation}
Comparing perturbed runs to the baseline calculation described in Section~\ref{sec:s3d} 
(again with a fixed $5$ns timestep and the 6,4-RK time integration method), we obtain 
a sensitivity profile for the various work arrays. 
Figure~\ref{fig:senscope} indicates the difference in the maximum
temperature in the simulation domain at a fixed time near the end of the ignition delay when the various
quantities are subjected individually to a one-time perturbation.  The
perturbation is applied to the output work array at the gridpoint where the
temperature is maximum by multiplying the value by $10^4$ immediately
after the value is calculated during the first substage function
evaluation for the timestep beginning at $t=5.5\mu\mathrm{s}$. We
observe, as many others have previously (e.g., \cite{FangPRG14}), that such error injection
can result in different categories of behavior:
\begin{enumerate}
\item The simulation fails in a detectable manner before completion,
frequently as soon as the perturbation is injected. This is typically
due to a non-realizable condition (e.g., temperature outside physical
bounds, the sum of the species mass fractions becoming much larger
than unity).
\item No detectable effect on the calculated ignition delay.
\item The calculation proceeds without apparent error to completion
and the calculated ignition delay is altered, with the size of the
error depending on the size of the perturbation earlier in the
calculation.
\end{enumerate}
While the first type may slow scientific progress due to
frequent restarts, it is the final type---the silent, undetectable
errors that alter the result of the calculation---that are the most
serious. 
Of the 93 kernel return values perturbed individually, for 74 of those
variables the calculation proceeds to completion. The remainder result
in simulation crashes (e.g., from out-of-bounds temperature extremes)
and are not shown in Figure~\ref{fig:senscope}. The error
in the temperature at the end of the solution ranged  from
70~K below the correct temperature) to 93~K above
the correct temperature; this corresponds to impacting the calculated ignition delay by more than
5\%. While it is difficult to make generalizations, perturbations that
increased the reaction rate involving known ignition promoters for this
mechanism ($O$, $OH$, $H$ ) resulted in a significant temperature
increase (hence, shorter ignition delay). Conversely, perturbations
that increased the transport rates and hence hindered the buildup of
radicals lead to a decrease in temperature (hence, longer ignition
delay).  The perturbation that increased the source term for the
continuity equation led to a decrease in temperature and is indicated
in green in Figure~\ref{fig:senscope} and will be considered in detail
in the following section as representative of the error injections
that led to silent data corruption.

\subsection{Solution after injection of perturbation} 
\begin{figure}[htb] \centering
  \includegraphics[width=0.8\columnwidth]{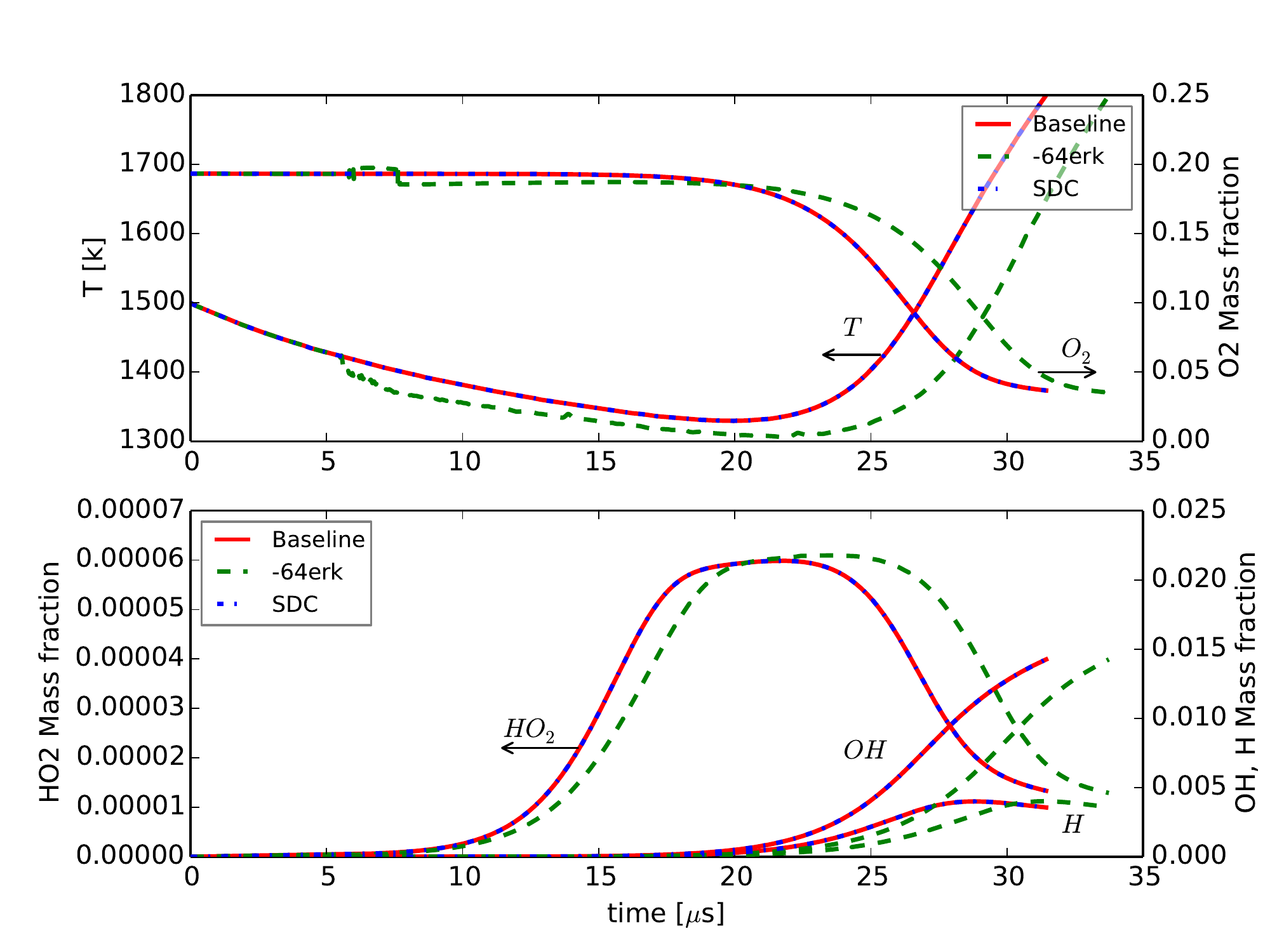}
  \caption{Effect of perturbation of continuity equation
    source term on solution
temporal evolution using 6,4-RK and SDC integration
schemes; temporal plots extracted at fixed spatial location where
error is injected. Notably, the SDC solution is indistinguishable
from the baseline solution while the Runge-Kutta solution is 
silently and significantly corrupted.}
  \label{fig:pertinjts}
\end{figure}

\begin{figure}[htb] \centering
  \includegraphics[width=0.75\columnwidth]{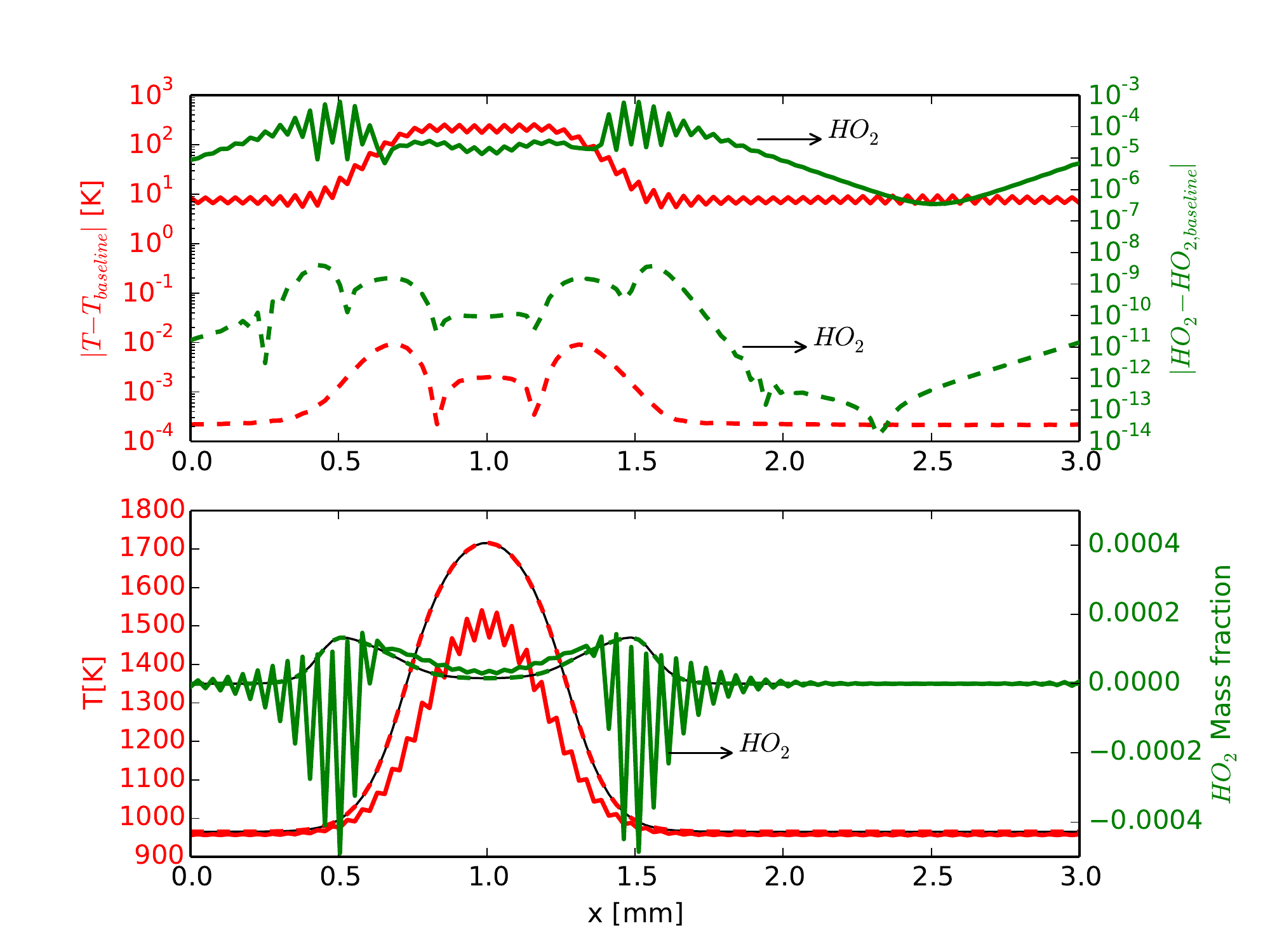}
    \caption{Effect of perturbation of continuity equation source term on solution
using 6,4-RK and SDC integration schemes at time of baseline
case ignition. Solid lines are the 6,4-RK solution and dashed lines
are the SDC solution for Temperature (red) and $HO_2$ mass fraction
(green). The upper plot shows the difference between the computed
solution and the baseline while the lower plot shows the computed
solution alongside the baseline (in solid black).}
  \label{fig:errorinjx}
\end{figure}

Modifying the term that forms the density time derivative in the
RHS evaluation, that is:
\begin{equation}
  \label{eq:dcontdx}
  \frac{\partial \rho}{\partial t} = \frac{\partial (-\rho u)}{\partial  x},
\end{equation}
results in a greater than $5\%$ increase in the eventual predicted ignition delay
and a significant change in the temperature at the end of the baseline
ignition delay as highlighted in Figure~\ref{fig:senscope} using the
6,4-RK integration method. Figure~\ref{fig:pertinjts} shows the temporal evolution
of the solution for temperature and key species for the baseline,
unperturbed case, for the 6,4-RK integration and for SDC
integration. The SDC integration is performed using three Gauss-Lobatto quadrature nodes and
four correction sweeps. 
Figure~\ref{fig:errorinjx} compares the spatial profiles at two different times,
the timestep after the error is injected and the timestep when the
baseline case reaches the ignition criterion.  
The perturbation grows
over time after the injection (at $t=5.5\mu\mathrm{s}$).  In keeping
with the silent nature of the corruption, by inspection of the portion
of the time history after the fault injection it is difficult to tell
that an error has occurred. Similarly, while it is obvious from looking
at the spatial profiles at later times in Figure~\ref{fig:errorinjx}
that the solution is contaminated by ringing, it is not clear how to
distinguish this from under-resolved physics~\cite{BrownM95}. Conversely, the solution
traces obtained when using SDC with a fixed number of iterations are
indistinguishable from the baseline, unperturbed case. This is an empirical
demonstration of the tendency of the  SDC iterations to recover from soft errors that
result in silently corrupted data when using the traditional
integration algorithm.  

\begin{figure}[htb] \centering
  \includegraphics[width=0.75\columnwidth]{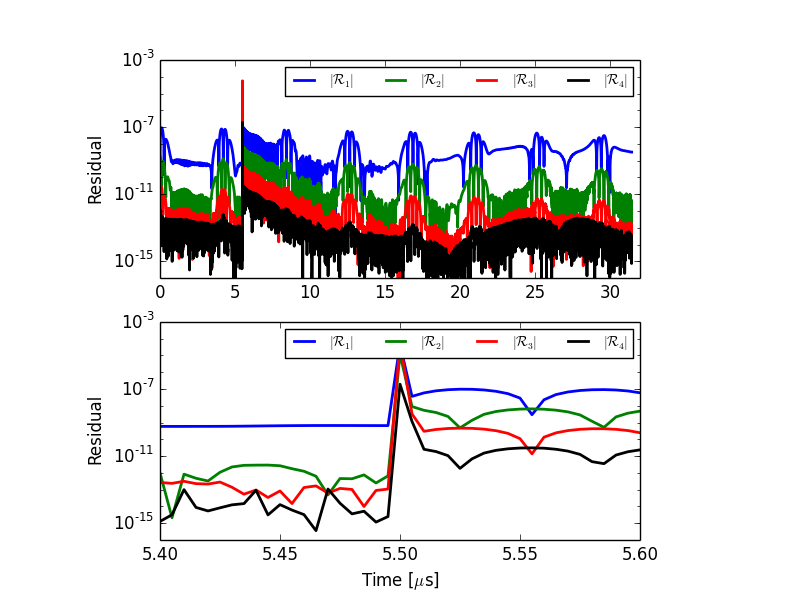}
  \caption{SDC residual response to soft perturbation for 4 correction
  sweeps given by $\mathcal{R}_1$--$\mathcal{R}_4$. Lower plot is a
  zoom in on the region of the fault injection; note that the fault is
  clearly evident by examining the residual. }
  \label{fig:sdcperturb}
\end{figure}

In Figure~\ref{fig:sdcperturb}, 
the residual as given in Equation~\ref{eq:residual} is shown over time; the 
curves shown for $|\mathcal{R}_k|$ are the magnitude of the residual for the
$k^\mathrm{th}$ correction iteration.  There is one value per time step plotted
obtained at the end of the time step; the lower portion of the figure is an
enlargement of the upper portion. We observe that the error injection can be
detected by monitoring the residual, which increases sharply when
the error is injected. In this experiment the number of SDC correction
iterations is held constant. While this is sufficient to damp the error
injected to the point
where the solution is not qualitatively deteriorated, the residual at the final
iteration has not reached its final value prior to the error injection. It is
several timesteps later that the residual after the final timestep reaches
approximately the same magnitude as the final residual prior to the error
injection. In the next section we will look at the response of a linear problem
to shed more light on how further SDC iterations reduce the error in a 
contaminated solution.

\clearpage
\subsection{Response of linear problem to perturbation}
\begin{figure}[htb] \centering
  \includegraphics[width=0.8\columnwidth]{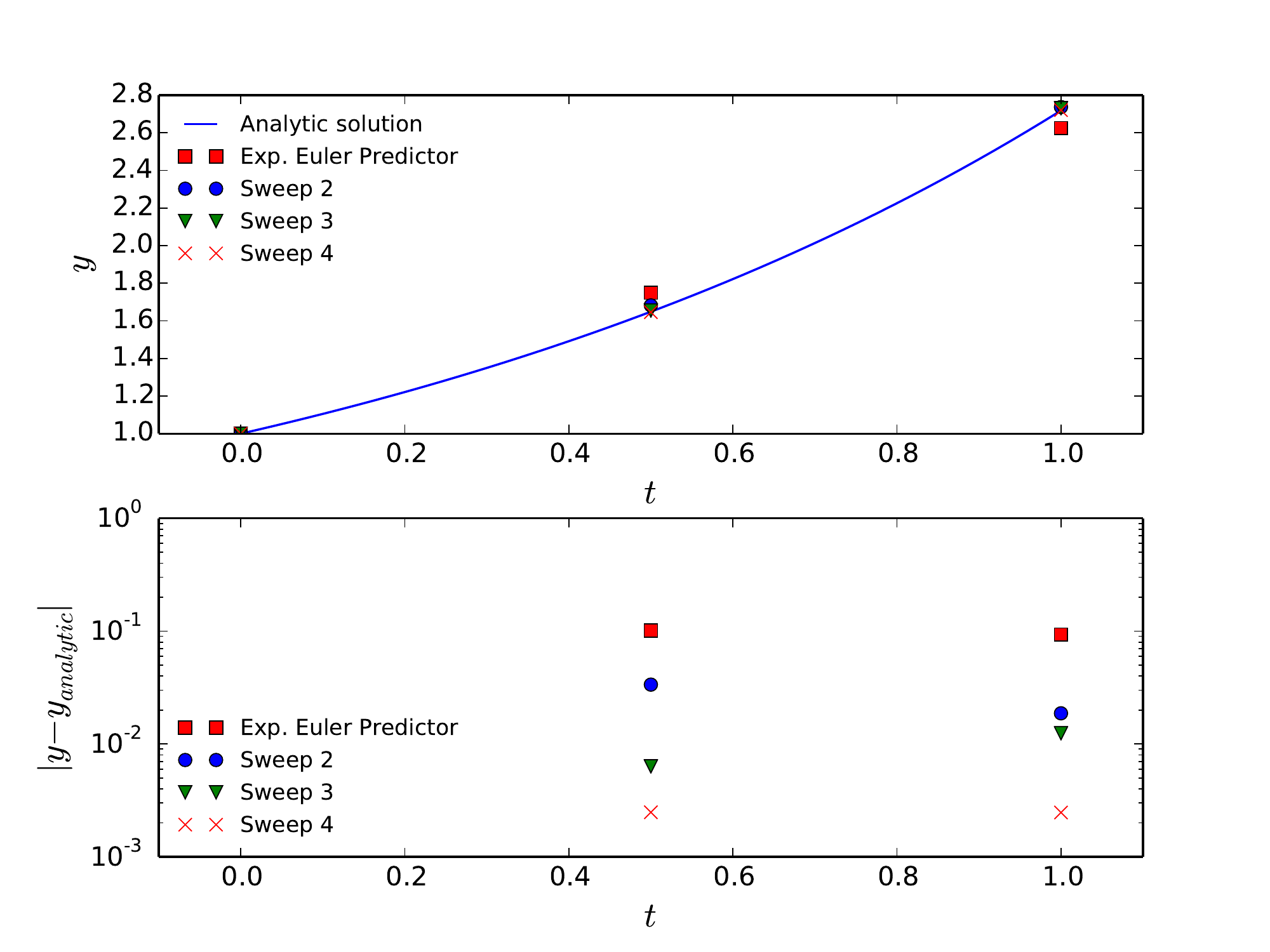}
  \caption{SDC iteration behavior for linear problem ($y'=Ay$)}
  \label{fig:linearSDC}
\end{figure}

\begin{figure}[htb] \centering
  \includegraphics[width=0.8\columnwidth]{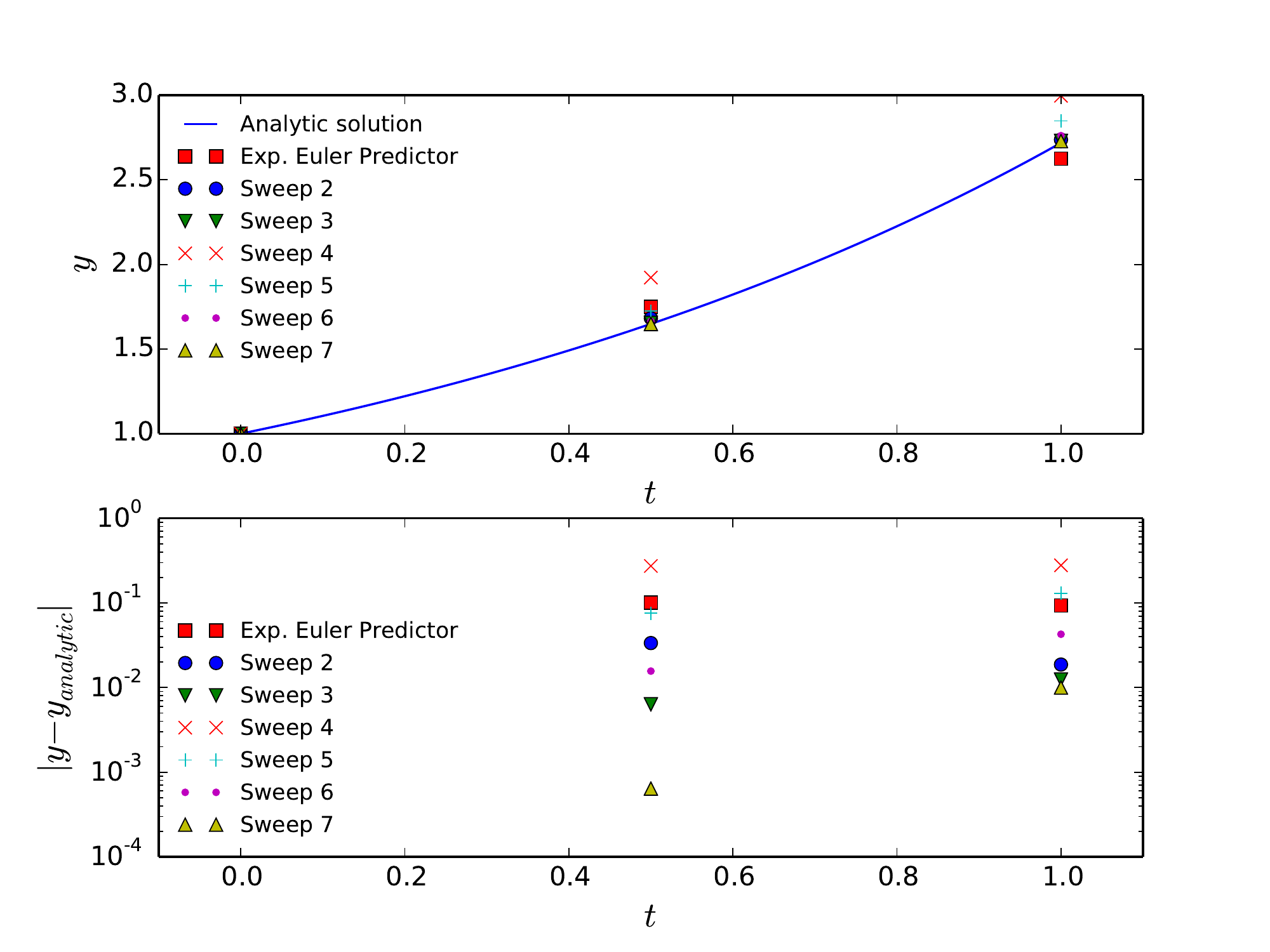}
  \caption{Effect of perturbation of linear problem on SDC iteration convergence. Error
  is injected during Sweep 3 which results in an error larger than the initial predictor but is
  then damped by sweeps 5 and 6.}
  \label{fig:linearSDCperturb}
\end{figure}

\begin{figure}[htb] \centering
  \includegraphics[width=0.8\columnwidth]{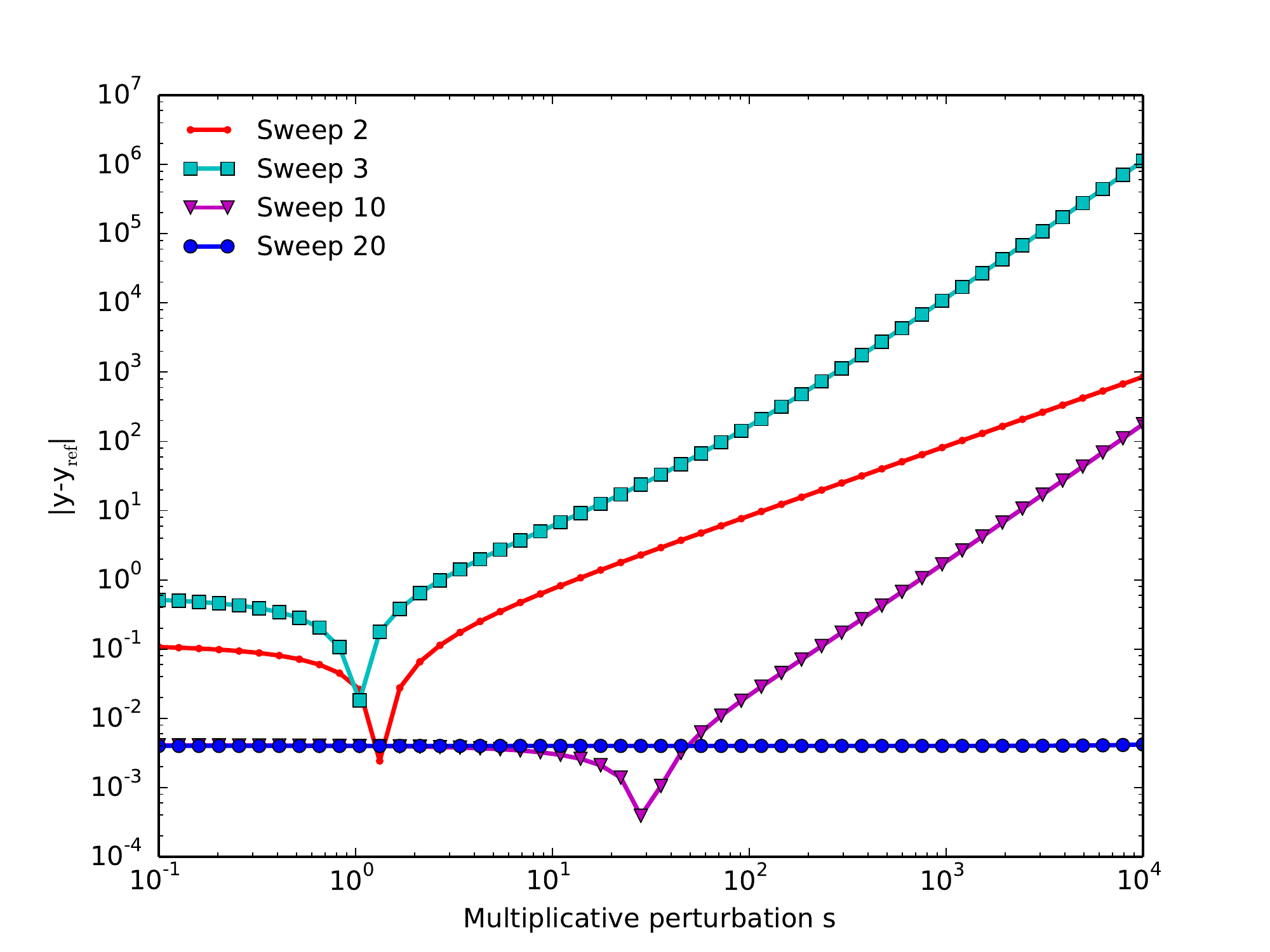}
  \caption{Effect of perturbation magnitude on SDC convergence rate. Reference solution $y_\mathrm{ref}$ is 
  analytic solution with $s=1$.}
  \label{fig:linearSDCdamping}
\end{figure}

In Figures~\ref{fig:linearSDC}~and~\ref{fig:linearSDCperturb}, a similar experiment is performed
on the linear test problem
\begin{equation}
  \label{eq:110}
  y'(t) = y(t) \qquad y(0) = 1
\end{equation}
over the interval $[0,1]$. 
Three quadrature nodes are used and four correction sweeps, including the initial explicit-Euler predictor,
are performed, giving a formally fourth-order method. The
baseline behavior is shown in Figure~\ref{fig:linearSDC} for
comparison to the perturbed solution in
Figure~\ref{fig:linearSDCperturb}. 
A perturbation to the solution is introduced 
by using $y' = (1.5)y$ for the derivative evaluation 
during the third SDC sweep at the second quadrature
node. 
For the unperturbed case, the solution error decreases
monotonically with iteration count as seen in
Figure~\ref{fig:linearSDC}. However, when the error is injected during
the third sweep, we see the error jump up again to near the error in
the initial predictor (Sweep 4 in
Figure~\ref{fig:linearSDCperturb}). After subsequent sweeps the error
is reduced until after Sweep 7 the error in the solution is less than
before the error is injected. 
Figure~\ref{fig:linearSDCdamping} demonstrates that the error damping is
geometric for a wide range of perturbation magnitudes. 
The horizontal axis in Figure~\ref{fig:linearSDCdamping}
corresponds to the size of the multiplicative perturbation to the 
derivative computation  $y'=sy$. We find that across a wide range of
$s$, both larger and smaller than unity, the error is damped with
a consistent ratio.
Also of note in Figure~\ref{fig:linearSDCdamping}, we look at continuing
the SDC iterations beyond the number of passes necessary for convergence
of the reference solution. Even for large perturbations that
result in errors several orders of magnitude larger than the reference solution converged
error, the converged solution remains the same. This feature of
SDC---the ability to recover from such large excursions from the true
solution---leads to its natural resilience. 

\subsection{Response to multiple errors}
\label{sec:multiple}

In this section we conduct an experiment to assess the potential of
the SDC iterations to  recover from soft errors in a more realistic
scenario. We use the baseline test case described above, running the
simulation until a fixed time $t_\mathrm{ign}$. We then extract the
maximum temperature in the domain as a global measure of the
simulation result. We set up our fault injection framework to inject
bit flips into random bits in the return values of randomly selected
kernels at random times, as discussed in
Section~\ref{sec:softerrinj}. Specifically, we injected 1 fault every 5580 calls to
the error injection callback per rank; this corresponds to approximately 1 fault
every 10 timesteps using the baseline RK time advance algorithm
without error injection. Within this window, each process (MPI rank)
chooses a random location where the fault will be injected. At the
start of the fault injection window, each rank initializes a counter
zero and chooses a random number in the range $(0,5580)$ to be the
fault call. The counter is incremented each time the error injection
callback is executed and when it equals the fault call a random bit
within the valid range of the argument pointer is flipped. The counter
continues to increment with successive calls, but without error
injection, until it reaches the window size when it is reset and a new
fault call count is chosen for the next window. For this 1-D
calculation 5 MPI ranks were used. 

When faults are injected randomly across the variable array, there is
the potential that some faults will result in immediate crashes of
the program as identified as the first type in Section~\ref{sec:sens}; i.e., flipping the sign bit of major variables or
changing the most significant bit in the exponent. These types of faults will cause the
program to experience an unrecoverable error that is
easily detectable. The test code solves a transport equation for
total energy and computes temperature by using a Newton search to
solve Equation~\ref{eq:eofT}.
hence, a bounds check on the temperature is likely to pick up
out-of-bounds issues on any of the variables that participate in
Equation~\ref{eq:eofT}. The code historically monitors the temperature
range during the solution of Equation~\ref{eq:eofT} and terminates if
it goes out of bounds. In order to allow the simulation to continue
without a full restart, we cache the solution vector at the start of
every outer timestep and allow the simulation to restart from that
point rather than terminating and restarting from a save-file.

As well, to deal with the final type of errors (those leading to
silent corruption), we modified the SDC algorithm to monitor its convergence
through the reduction in the residual. We propose a strategy for mitigating soft errors---hardware
introduced faults that are stochastic and transient in nature---
based on monitoring the behavior of the SDC correction through the
residual to identify when a soft error has occurred and continuing
iteration until the residual drops to the prescribed tolerance. In the
case of non-recoverable errors detected during the correction iterations
we restart the timestep.  For each correction iteration
(after the first) we compute:
\begin{eqnarray}
  \label{eq:16}
  \mathcal{R}_1 = \frac{\max |\vec{R}_n|  }{ \max |\vec{R}_1| } \\
  \mathcal{R}_{n-1} = \frac{ \max |\vec{R}_n| }{\max |\vec{R}_{n-1}|  }
\end{eqnarray}
and continue the correction sweeps until $\mathcal{R}_1 < 10^{-5}$ and
$\mathcal{R}_{n-1} > 0.9$, that is, until the residual is small
compared to the residual from the first correction pass and is also
not changing significantly between successive correction passes. 
The tolerance values for $\mathcal{R}_1$ and $\mathcal{R}_{n-1} $
were chosen to be consistent with the
ratios found in the baseline case without fault injection at the end
of the SDC iterations. The maximum number of correction passes is
limited to 8, after which the timestep is accepted. In practice, only
a few timesteps encountered this limit.

\begin{figure}[htb] \centering
  \includegraphics[width=0.8\columnwidth]{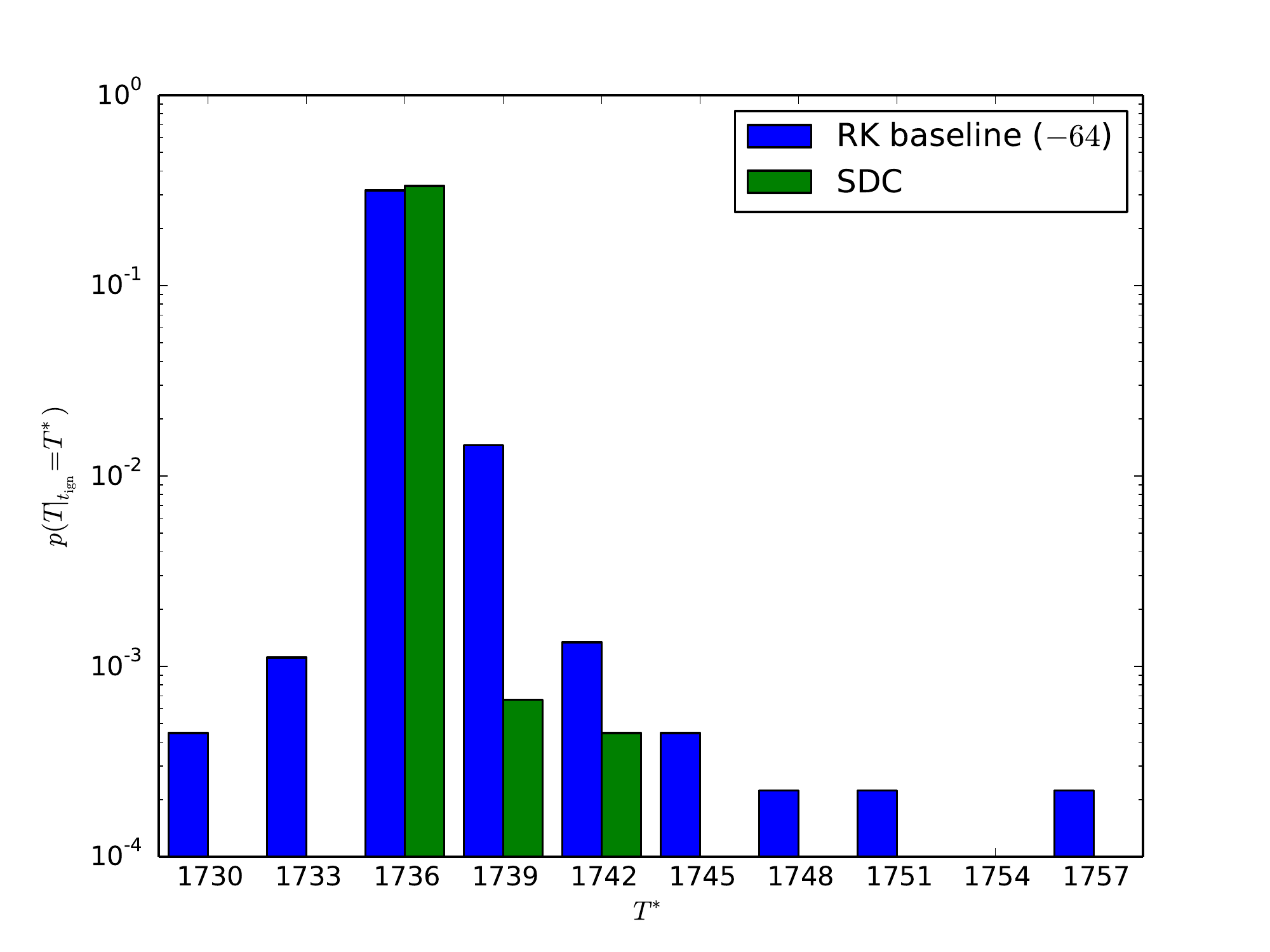}
  \caption{Distribution of temperature at end of calculation with error injection using baseline Runge-Kutta integration and SDC approach of same the order }
  \label{fig:errdist}
\end{figure}

We conducted 1500 independent runs using both the baseline RK time
integration and the proposed SDC method;  the distribution of the
temperature at the end of the calculation is shown in
Figure~\ref{fig:errdist} and Table~\ref{tab:errdisttab}.
\begin{table}[h!]
  \centering
  \begin{tabular}{l | c c c c c}
    \hline
 & Mean & Minimum & Maximum & Span & Variance \\ 
\hline
 RK & 1737.32  & 1728.92 & 1758.82 & 29.90 & 0.95  \\ 
 SDC & 1737.30  & 1736.70 & 1743.69 & 7.00 & 0.04  \\ 
  \end{tabular}
  \caption{Variance in temperature at end of calculation with error
 injection using baseline Runge-Kutta integration and SDC
 approach of the same order. Result from both methods without error injection is 1737.25.} 
 \label{tab:errdisttab}
\end{table}
The data in the table demonstrates that the temperature values using
SDC are significantly more clustered near the reference value than
those computed using RK.  There are some occurrences
where the error introduced is sufficiently large that 
maximum SDC iteration limit is reached 
before without fully damping the error, which
accounts for the non-zero variance in the
sample of the SDC solutions. Despite this, the width of the
distribution is far narrower than the corresponding baseline
distribution. In a production environment, two alternatives to narrow
the distribution further are available: more SDC iterations could be
allowed, or the timestep could be restarted if the iteration limit is
reached. For this test, the rate of error injection is significantly
magnified from realistic error rates, so either option is
likely acceptable with minimal computational cost under realistic
error rates for a target platform. This is meant to be illustrative:
given the uncertainty in the error rates for future architectures, we
demonstrate that the simulation can make progress and the effect of
those errors mitigated, but it is difficult to assess computational
cost without knowing what the error rates are. This is left for
future work as more realistic predictions and measurements of soft
error rates on extreme scale architectures become
available. Satisfyingly, the resilient form of SDC does not add extra
cost beyond a general formulation when there are no hardware faults.
 In the presence of
extreme error rates, the algorithm still makes progress, with the
vast majority of runs resulting in no silent data corruption and a
clear path to including the remaining outliers available. 
\section{Conclusion}
Natural extensions to a generic SDC algorithm have been proposed that are
demonstrated to provide improved algorithmic resilience. It is shown that, in the face of a
single transient error, continued SDC iterations beyond those normally
required provide a viable approach to error recovery. In the case of
 elevated rates of stochastic errors, 
 the algorithm can still make progress. In addition, although it is not explored here,
the method provides a mechanism for detecting stuck bit errors that 
could potentially be used to trigger restarting the affected timestep
using different memory for the work arrays. When no errors are introduced, the
suggested formulation reverts to a generic SDC algorithm so there is no
significant cost penalty for the modifications. The formulation is
predicated on the ability to protect the integrity of the solution state
between successive timesteps, as well as the program control flow.
However, the work arrays used by the application code during a timestep
that typically comprise the majority of the memory usage can be exposed
to significant error rates. This provides an opportunity for a savings,
where the need for error correction is potentially reduced without
resorting to measures such as redundant calculation that increases
computational cost irrespective of the actual error rate realized. As
such, the method is a way for application developers to design for
potential increased soft error rates on future hardware without the
penalty of degraded performance on less error-prone architectures.

\section*{Acknowledgments}
This material is based upon work supported by the U.S. Department of
Energy, Office of Science, Office of Advanced Scientific Computing
Research.
Work at LBNL was supported the Co-Design Program of the U.S. DOE Office
of Advanced Scientific Computing Research under contract DE-AC02005CH11231.

\bibliographystyle{siam}
\bibliography{../writeup/bib/pid_sdc.bib}

\begin{thebibliography}{10}

\bibitem{Borkar99}
{\sc S.~Borkar}, {\em Design challenges of technology scaling}, IEEE Micro, 19
  (1999), pp.~23--29.

\bibitem{bourliouxLaytonMinion:2003}
{\sc A.~Bourlioux, A.~T. Layton, and M.~L. Minion}, {\em Higher-order
  multi-implicit spectral deferred correction methods for problems of reacting
  flow}, Journal of Computational Physics, 189 (2003), pp.~351--376.

\bibitem{BrownM95}
{\sc D.~L. Brown and M.~Minion}, {\em Performance of under-resolved
  two-dimensional incompressible flow simulations}, Journal of Computational
  Physics, 121 (1995).

\bibitem{s3d}
{\sc J.~H. Chen, A.~Choudhary, B.~de~Supinski, M.~DeVries, E.~R. Hawkes,
  S.~Klasky, W.~K. Liao, K.~L. Ma, J.~Mellor-Crummey, N.~Podhorszki,
  R.~Sankaran, S.~Shende, and C.~S. Yoo}, {\em Terascale direct numerical
  simulations of turbulent combustion using {S3D}}, Comput. Sci. Disc., 2
  (2009), p.~015001.

\bibitem{ChenBCP13}
{\sc S.~Chen, G.~Bronevetsky, M.~Casas-Guix, and L.~Peng}, {\em Comprehensive
  algorithmic resilience for numeric applications}, Tech. Report
  LLNL-CONF-618412, Lawrence Livermore National Laboratory (LLNL), Livermore,
  CA, 2013.

\bibitem{Constantinescu02}
{\sc C.~Constantinescu}, {\em Impact of deep submicron technology on
  dependability of vlsi circuits}, in Dependable Systems and Networks, 2002.
  DSN 2002. Proceedings. International Conference on, 2002, pp.~205--209.

\bibitem{DegalahalRVXI03}
{\sc V.~Degalahal, R.~Ramanarayanan, N.~Vijaykrishnan, Y.~Xie, and M.J. Irwin},
  {\em The effect of threshold voltages on the soft error rate [memory and
  logic circuits]}, in Quality Electronic Design, 2004. Proceedings. 5th
  International Symposium on, 2004, pp.~503--508.

\bibitem{DonzisA14}
{\sc D.~A. Donzis and K.~Aditya}, {\em Asynchronous finite-difference schemes
  for partial differential equations}, Journal of Computational Physics, 274
  (2014).

\bibitem{DuttGR00}
{\sc A.~Dutt, L.~Greengard, and V.~Rokhlin}, {\em Spectral deferred correction
  methods for ordinary differential equations}, BIT Numerical Mathematics, 40
  (2000), pp.~241--266.

\bibitem{EchekkiC03}
{\sc T.~Echekki and J.~H. Chen}, {\em Direct numerical simulation of
  autoignition in non-homogeneous hydrogen-air mixtures}, Combustion and Flame,
  134 (2003), pp.~169--191.

\bibitem{ElliottMSW13}
{\sc J.~Elliott, F.~Mueller, M.~Stoyanov, and C.~Webster}, {\em Quantifying the
  impact of single bit flips on floating point arithmetic}, tech. report, Tech.
  Rep. ORNL/TM-2013/282, Oak Ridge National Laboratory, One Bethel Valley Road,
  Oak Ridge, TN, 2013. 6, 9, 2013.

\bibitem{EmmettM12}
{\sc M.~Emmett and M.~L. Minion}, {\em Towards an efficient parallel in time
  method for partial differential equations}, Communications in Applied
  Mathematics and Computational Science, 7 (2012), pp.~105--132.

\bibitem{FangPRG14}
{\sc B.~Fang, K.~Pattabiraman, M.~Ripeanu, and S.~Gurumurthi}, {\em Evaluating
  the error resilience of parallel programs}, in The 4th fault tolerance for
  {HPC} at e{X}treme scale (FTXS), Atlanta, Georgia, June 2014.

\bibitem{GroutGKBBGC12}
{\sc R.W. Grout, A.~Gruber, H.~Kolla, P.-T. Bremer, J.C. Bennett, A.~Gyulassy,
  and J.H. Chen}, {\em {A direct numerical simulation study of turbulence and
  flame structure in a round jet in cross-flow}}, Journal of Fluid Mechanics,
  706 (2012), pp.~351--383.

\bibitem{GroutGYC11}
{\sc R.W. Grout, A.~Gruber, C.S. Yoo, and J.H. Chen}, {\em {Direct numerical
  simulation of flame stabilization downstream of a transverse fuel jet in
  cross-flow}}, Proceedings of the Combustion Institute, 33 (2011),
  pp.~1629--1637.

\bibitem{GruberSHC10}
{\sc A.~Gruber, R.~Sankaran, E.~R. Hawkes, and J.~H. Chen}, {\em Turbulent
  flame--wall interaction: a direct numerical simulation study}, Journal of
  Fluid Mechanics, 658 (2010), pp.~5--32.

\bibitem{HawkesC05}
{\sc E.~R. Hawkes and J.~H. Chen}, {\em Evaluation of models for flame stretch
  due to curvature in the thin reaction zones regime}, in Proceedings 30th
  International Symposium on Combustion, The Combustion Institute, 2005,
  pp.~647--655.

\bibitem{HawkesSSC07}
{\sc E.~R. Hawkes, R.~Sankaran, J.~C. Sutherland, and J.~H. Chen}, {\em Scalar
  mixing in direct numerical simulations of temporally evolving plane jet
  flames with skeletal co/h2 kinetics}, in Proceedings 31th International
  Symposium on Combustion, The Combustion Institute, 2007, pp.~1633--1640.

\bibitem{HoemmenH11}
{\sc M.~Hoemmen and M.~A. Heroux}, {\em Fault-tolerant iterative methods via
  selective reliability}, in Proceedings of the 2011 International Conference
  for High Performance Computing, Networking, Storage and Analysis (SC),
  vol.~3, IEEE Computer Society, 2011.

\bibitem{HwangSS12}
{\sc Andy~A. Hwang, Ioan~A. Stefanovici, and Bianca Schroeder}, {\em Cosmic
  rays don't strike twice: Understanding the nature of dram errors and the
  implications for system design}, in {ASPLOS XVII} Proceedings of the
  seventeenth international conference on Architectural Support for Programming
  Languages and Operating Systems, ACM, 2012, pp.~111--122.

\bibitem{KeeRMM96}
{\sc R.~J. Kee, R.~M. Ruply, E.~Meeks, and J.~A. Miller}, {\em Chemkin-{III}: A
  {FORTRAN} chemical kinetics package for the analysis of gas-phase chemical
  and plasma kinetics}, Technical Report SAND96-8216, Sandia National
  Laboratories, Livermore, 1996.

\bibitem{KennedyCL00}
{\sc C.~A. Kennedy, M.~H. Carpenter, and R.~M. Lewis}, {\em Low-storage,
  explicit {R}unge-{K}utta schemes for the compressible {N}avier-{S}tokes
  equations}, Applied Numerical Mathematics, 35 (2000), pp.~177--219.

\bibitem{LaytonM04}
{\sc A.~T. Layton and M.~L. Minion}, {\em Conservative multi-implicit spectral
  deferred correction methods for reacting gas dynamics}, Journal of
  Computational Physics, 194 (2004), pp.~697--715.

\bibitem{LaytonM05}
\leavevmode\vrule height 2pt depth -1.6pt width 23pt, {\em Implications of the
  choice of quadrature nodes for {P}icard integral deferred correction
  methods}, BIT, 45 (2005), pp.~341--373.

\bibitem{LiZKD04}
{\sc J.~Li, Z.~Zhao, A.~Kazakov, and F.~L. Dryer}, {\em An updated
  comprehensive kinetic model of hydrogen combustion}, International Journal of
  Chemical Kinetics, 36 (2004), pp.~566--575.
\newblock DOI 10.1002/kin.20026.

\bibitem{MayoAR12}
{\sc J.~Mayo, R.~Armstrong, and J.~Ray}, {\em Efficient, broadly applicable
  silent-error tolerance for extreme-scale resilience}, Tech. Report
  SAND2012-8131, Sandia National Laboratories, 2012.

\bibitem{Minion03}
{\sc M.~L. Michael}, {\em Semi-implicit spectral deferred correction methods
  for ordinary differential equations}, Communications in Mathematical
  Sciences, 1 (2003), pp.~471--500.

\bibitem{NonakaBDGAM12}
{\sc A.~Nonaka, J.~B. Bell, M.~S. Day, C.~Gilet, A.~S. Almgren, and M.~L.
  Minion}, {\em {A deferred correction coupling strategy for low Mach number
  flow with complex chemistry}}, Combust. Theory \& Model.,  (2012), pp.~1--36.

\bibitem{SankaranHCLL07}
{\sc R.~Sankaran, E.~R. Hawkes, J.~H. Chen, T.~Lu, and C.~K. Law}, {\em
  Structure of a spatially developing turbulent lean methane-air bunsen flame},
  in Proceedings 31th International Symposium on Combustion, The Combustion
  Institute, 2007, pp.~1291--1298.

\bibitem{SankaranIHC05}
{\sc R.~Sankaran, H.~G. Im, E.~R. Hawkes, and J.~H. Chen}, {\em The effects of
  non-uniform temperature distribution on the ignition of a lean homogeneous
  hydrogen-air mixture}, in Proceedings 30th International Symposium on
  Combustion, The Combustion Institute, 2005, pp.~875--882.

\bibitem{SchroederPW11}
{\sc B.~Schroeder, E.~Pinheiro, and W.-D. Weber}, {\em Dram errors in the wild:
  a large-scale field study}, Commun. ACM, 54 (2011), pp.~100--107.

\bibitem{SloanKB13}
{\sc J.~Sloan, R.~Kumar, and G.~Bronevetsky}, {\em An algorithmic approach to
  error localization and partial recomputation for low-overhead fault
  tolerance}, 2013 43rd Annual IEEE/IFIP International Conference on Dependable
  Systems and Networks (DSN), 0 (2013), pp.~1--12.

\bibitem{SpaffordMVCGS09}
{\sc K.~Spafford, J.~Meredith, J.S. Vetter, J.~Chen, R.~Grout, and
  R.~Sankaran}, {\em Accelerating {S3D}: A {GPGPU} case study}, in Seventh
  International Workshop on Algorithms, Models, and Tools for Parallel
  Computing on Heterogeneous Platforms (HeteroPar 2009), Delft, The
  Netherlands, 2009.

\bibitem{SridharanL12}
{\sc V.~Sridharan and D.~Liberty}, {\em A study of dram failures in the field},
  in Proceedings of the International Conference on High Performance Computing,
  Networking, Storage and Analysis, SC '12, Los Alamitos, CA, USA, 2012, IEEE
  Computer Society Press, pp.~76:1--76:11.

\bibitem{SridharanSDBG13}
{\sc V.~Sridharan, J.~Stearley, N.~DeBardeleben, S.~Blanchard, and
  S.~Gurumurthi}, {\em Feng shui of supercomputer memory: positional effects in
  dram and sram faults}, in SC, William Gropp and Satoshi Matsuoka, eds., ACM,
  2013, p.~22.

\bibitem{StoyanovW13}
{\sc M.~K. Stoyanov and C.~G. Webster}, {\em Numerical analysis of fixed point
  algorithms in the presence of hardware faults}, tech. report, Oak Ridge
  National Laboratory (ORNL), 2013.

\bibitem{WeiTLP14}
{\sc J.~Wei, A.~Thomas, G.~Li, and K.~Pattabiraman}, {\em Quantifying the
  accuracy of high-level fault injection techniques for hardware faults}, in
  IEEE/IFIP International Conference onf Dependable Systems and Networks,
  Atlanta, Georgia, 2014.

\bibitem{YooSC09}
{\sc C.~S. Yoo, R.~Sankaran, and J.~H. Chen}, {\em Three-dimensional direct
  numerical simulation of a turbulent lifted hydrogen jet flame in heated
  coflow: Flame stabilization and structure}, Journal of Fluid Mechanics, 640
  (2009), pp.~453--481.

\end{thebibliography}
\end{document}